\documentstyle[11pt,newpasp,twoside,epsf,astrobib]{article}
\def\edcomment#1{\iffalse\marginpar{\raggedright\sl#1\/}\else\relax\fi}
\reversemarginpar

\begin{document}
\title{X-Ray Constraints on Dark Matter in Galaxy Clusters and Elliptical
Galaxies: A View from Chandra and XMM} 
\author{David A. Buote}
\affil{University of California, Irvine, 4129 Frederick Reines Hall,
Irvine, California, 92612-4575 U.S.A., buote@uci.edu}

\begin{abstract}
I review constraints on the radial density profiles and ellipticities
of the dark matter obtained from recent X-ray observations with {\sl
Chandra} and {\sl XMM} of elliptical galaxies and galaxy clusters and
discuss their implications, especially for the self-interacting dark
matter model.
\end{abstract}

\section{Introduction}

For many years X-ray astronomers have promised to obtain accurate
constraints on dark matter in clusters of galaxies and elliptical
galaxies. But because of the frustrating limitations of previous X-ray
telescopes, only for a very few objects -- notably M87 -- have precise
measurements been possible. It is really a great pleasure to give this
review because the promises made many years ago are finally being
realized in this wonderful era of X-ray astronomy, where the {\sl
Chandra} and {\sl XMM} observatories are operating so successfully.

{\sl Chandra} and {\sl XMM} have provided for the first time high
quality, spatially resolved spectra of the diffuse hot gas of galaxies
and clusters because their CCDs combine moderate resolution spectra
with very much improved spatial resolution and sensitivity. {\sl
Chandra} provides a more significant jump in spatial resolution while
XMM provides a more substantial boost in sensitivity. As a result of
these improved capabilities, accurate measurements of the gas
temperature as a function of radius exist for many clusters. These
measurements provide very interesting constraints on the DM.

Because most of the published results on X-ray studies of dark matter
(DM) using {\sl Chandra} and {\sl XMM} exist for clusters, in this
review I will emphasize the results obtained on the radial DM
distributions in clusters. My discussion will be divided up into
segments that address the mass distributions inside and outside of
cluster cores. I devote the remaining space to elliptical galaxies,
particularly NGC 720, where I will discuss X-ray constraints on the
ellipticity of DM halos.

\section{Galaxy Clusters}

In galaxy clusters the dominant baryonic component is that of the hot
gas, yet it contributes only 10-30 percent to the total mass. Clusters
are therefore ``DM-dominated'' and are excellent sites to study the
properties of the DM. In fact, in the context of the CDM model,
simulations by \citeN{dubi98a} suggest that clusters are DM-dominated
down to less than 1\% of the virial radius ($0.01r_{\rm vir}$), making
them especially attractive for studies of the cores of DM
halos. Another advantage of studying clusters is that there are a
variety of methods that can be used to probe their mass distributions
-- stellar/galaxy dynamics, gravitational lensing, and dynamics of the
hot (X-ray) gas. Each of these methods has certain advantages and
disadvantages. The X-ray method, which is the focus here, is primarily
hampered by the assumption of hydrostatic equilibrium and questions
about the thermodynamic state of the hot gas.

As for the assumption of hydrostatic equilibrium, provided one selects
clusters with regular morphologies, hydrodynamic simulations show that
the X-ray method is robust, even if the cluster is not in perfect
equilibrium \cite{tsai94a,evra96a,math99a}. Further support for
hydrostatic equilibrium is indicated by the generally good agreement
between cluster masses obtained from weak lensing and X-rays
\cite{alle02a}, though some disagreements with strong lensing remain
\cite{mach02,xue02}. 

Regarding the state of the hot gas, mass determinations using X-ray
data usually assume the gas is single-phase. Indeed, the {\sl Chandra}
and {\sl XMM} observations of clusters have justified that assumption
(e.g., \citeNP{pete01}; \citeNP{tamu01}). These observations have
shown that outside cluster cores the hot gas is single-phase. However,
within cluster cores the data are consistent with, but do not
necessarily require, a limited multiphase medium with a temperature
range much less than that expected from a standard multiphase cooling
flow. In a few of the cooler systems there is clear evidence for
multiphase gas in their cores (M87, Centaurus, NGC 5044). Although the
single-phase assumption certainly appears valid outside cluster cores,
the situation is less clear within the cores of cool clusters and
groups.

For a single-phase gas in hydrostatic equilibrium the calculation of
the radial mass distribution from X-ray data is fairly
straightforward. Typically, one assumes spherical symmetry and divides
up the X-ray image into a series of concentric, circular annuli. Then
coronal plasma models are fitted to the annular spectra to infer the
temperature and density of the gas as a function of radius. Often this
procedure is done by first deprojecting the data using an ``onion
peeling'' method pioneered by Andy Fabian and collaborators. Then
parameterized functions are fitted to the radial profiles of the gas
density and temperature to allow analytical calculation of the
derivatives in the hydrostatic equation. The effects of rotation and
magnetic fields are usually ignored but should be negligible
\cite{math03a}. Data from next-generation X-ray satellites should
provide the first interesting constraints on gas rotation.

\subsection{Outside the Core}

\begin{figure}
\plottwo{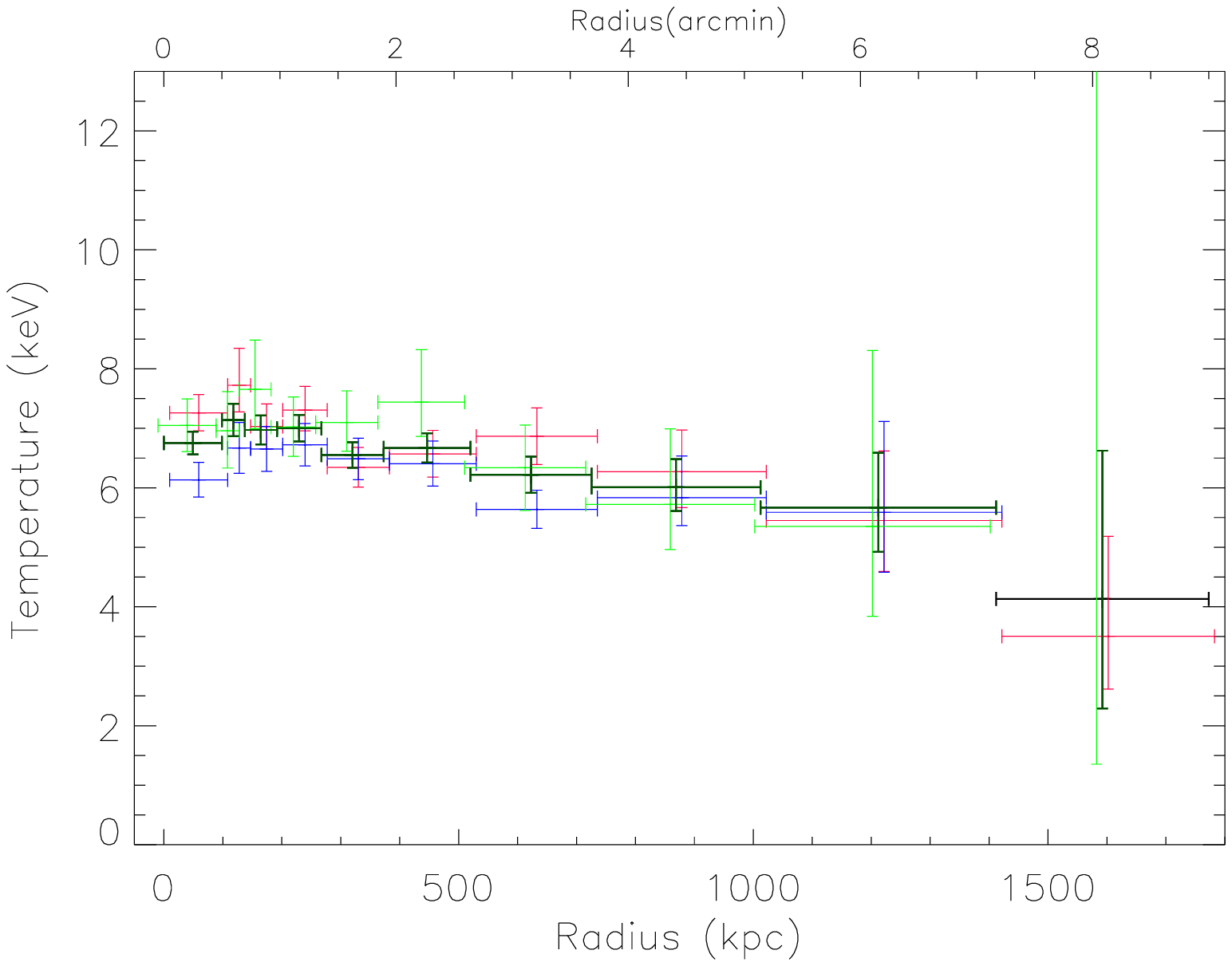}{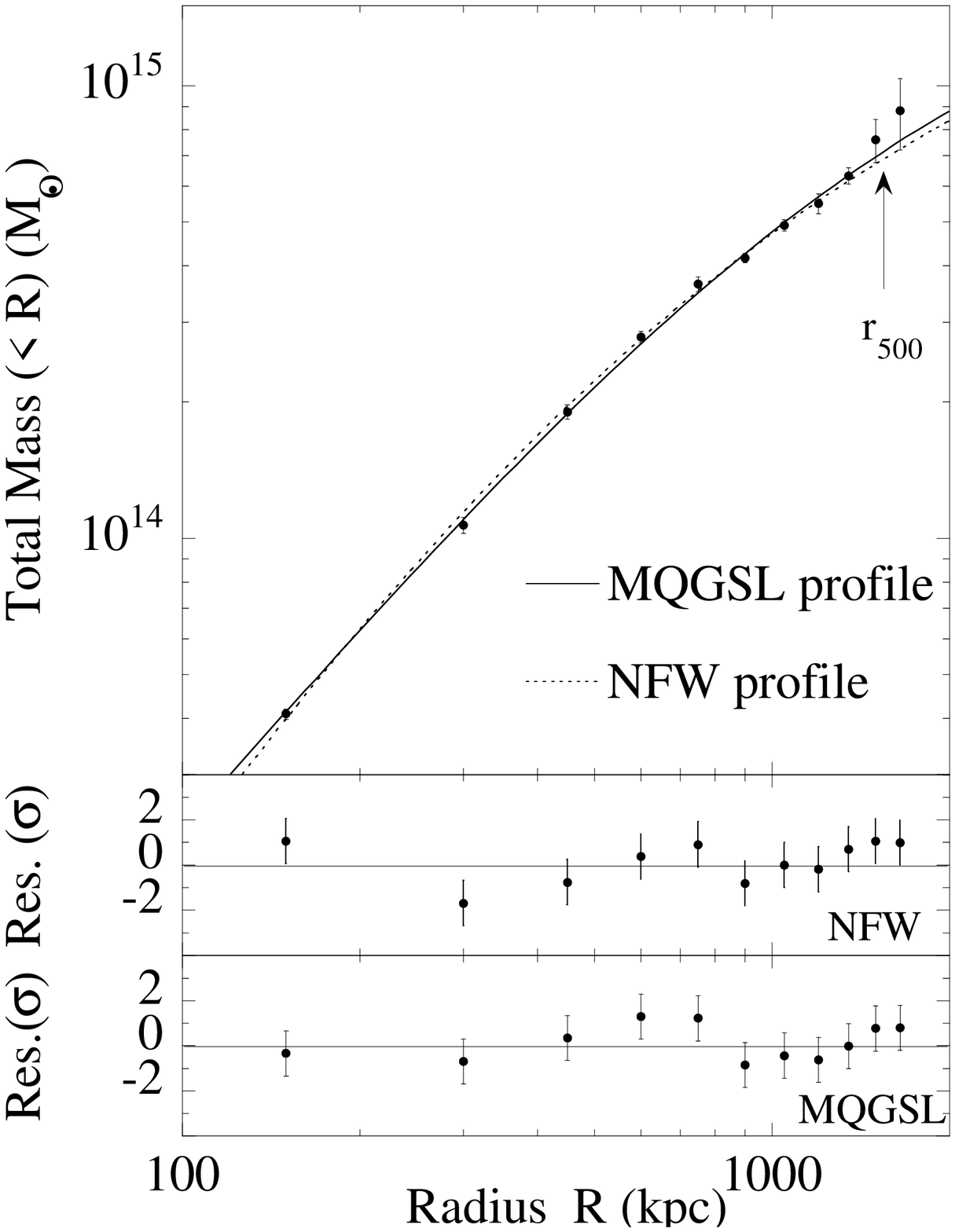}
\caption{\label{fig.a1413} From Pratt \& Arnaud's (2002) analysis of
the {\sl XMM} data 
of the bright galaxy cluster A1413. ({\it Left Panel}) Radial
temperature profile of the hot gas, and ({\it Right Panel}) the radial
mass profile.}
\end{figure}

Let us consider first the results obtained with {\sl Chandra} and {\sl
XMM} for the mass profiles of clusters outside their cores. Perhaps
the most impressive example is that of A1413 studied by
\citeN{prat02a}. This bright cluster has a very regular X-ray image,
the isophotes are moderately flattened, and the radial surface
brightness profile shows no structure other than a central enhancement
over a single $\beta$ model. These authors obtained an excellent
determination of the temperature profile between 100~kpc and 1.5~Mpc
(see Figure \ref{fig.a1413}). Pratt \& Arnaud determined the mass
profile by examining different parameterizations for the temperature
and density of the hot gas. Outside their central data point, they
found the gravitating mass profile is very precisely constrained and
is quite insensitive for the specific parameterizations of the gas
density and temperature. They find that the NFW and Moore et al.\
profiles provide good fits to the shape of the gravitating mass
profile from $\approx 0.1r_{\rm vir}$ out to $0.7r_{\rm vir}$ and give
reasonable values for the concentration parameter, $c=5.4\pm 0.2$ (for
NFW).

A1835 is another example of a generally relaxed
cluster. \citeN{schm01a} obtained good constraints on the mass profile
using {\sl Chandra} data. Their approach to constrain the mass is a
slight variation on the procedure mentioned above. They start with an
assumed mass profile and use the well-constrained gas density profile
to predict the radial gas temperature profile which is then fitted to
the data. Schmidt et al.\ find acceptable fits to their temperature
profile out to 1~Mpc $(\approx 0.4r_{\rm vir})$ using an NFW model and
obtain a reasonable value for the concentration $(c=4.0\pm 0.6)$. They
also find that their mass profile agrees with that obtained from a
strong lensing analysis.

\begin{figure}
\plottwo{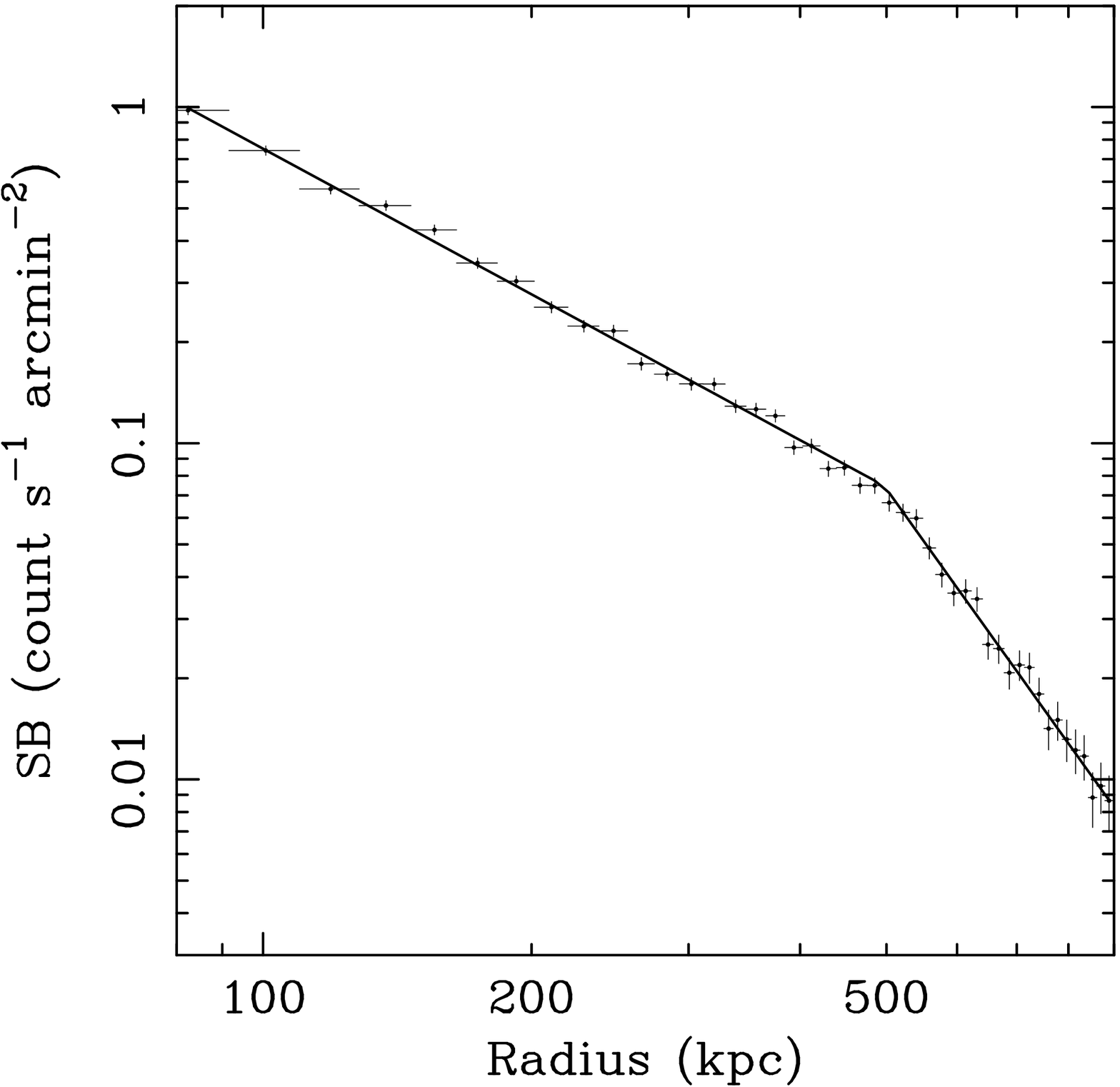}{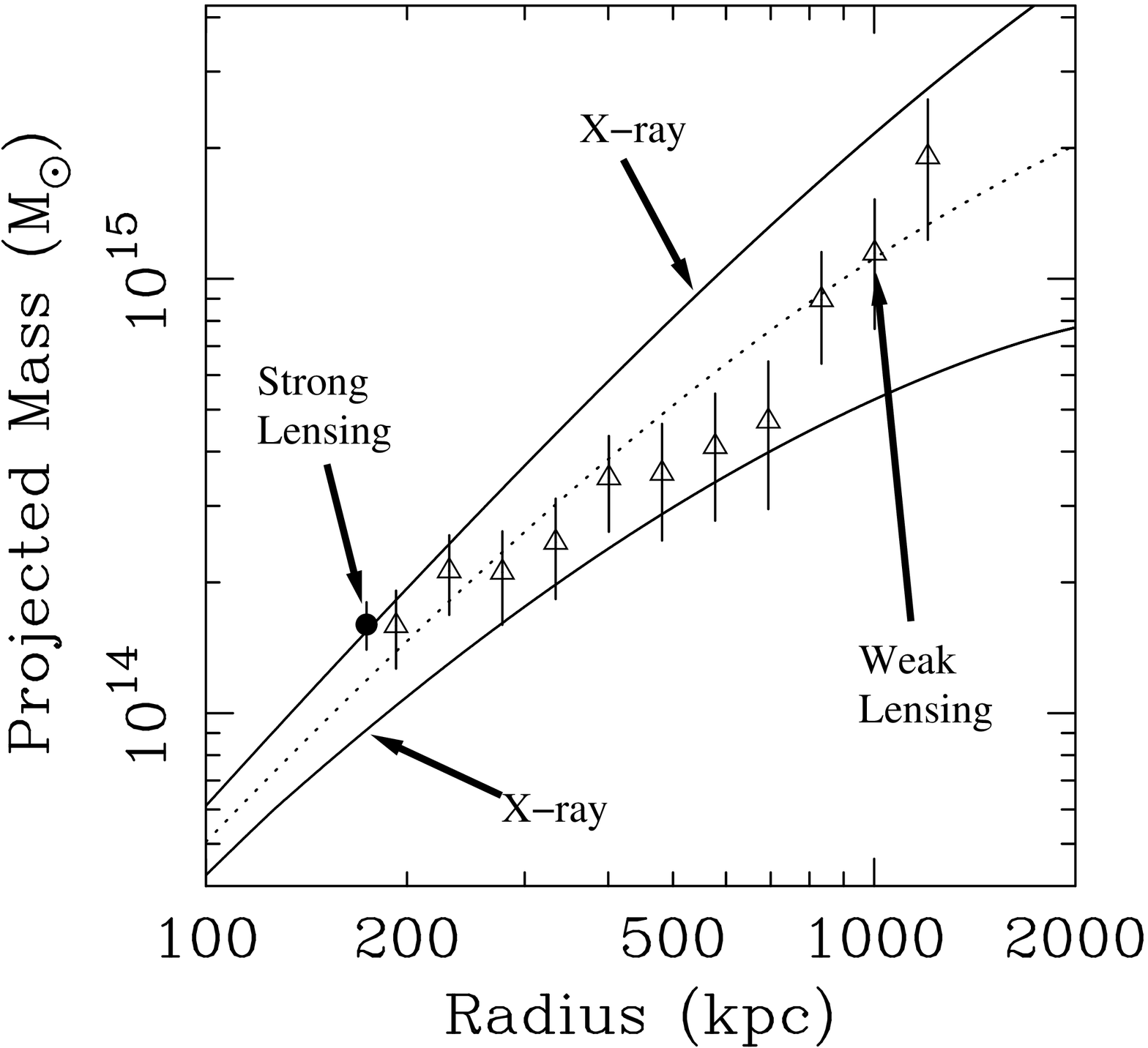}
\caption{\label{fig.a2390} From Allen, Ettori, \& Fabian's (2001)
analysis of the {\sl Chandra} data of the lensing galaxy cluster
A2390. ({\it Left Panel}) Radial surface profile of the hot gas
showing a peculiar break, and ({\it Right Panel}) the radial mass
profile showing the good agreement between the mass determined by both
strong and weak lensing and X-rays.}
\vspace{-0.65cm}
\end{figure}

{\sl Chandra} results for A2390 (Figure \ref{fig.a2390}) -- another
lensing cluster -- have been presented by \citeN{alle01d}. This
cluster is not as regular as those just discussed. Instead, there is
some substructure in the core and irregularities in the X-ray image on
$\approx 100$~kpc scales. The radial surface brightness profile even
shows a curious break near a radius of $0.5$~Mpc. Hence, this is
certainly a cluster where the assumption of hydrostatic equilibrium is
suspect. However, Allen et al.\ find that the NFW model is a good fit
to the data and gives a reasonable value for the concentration
$(c=3.3\pm 1.7)$. Very good agreement is also observed between the
X-ray--determined mass and that obtained from strong and weak lensing
studies. Thus, whatever departures from hydrostatic equilibrium exist
in this cluster do not appear to translate to large errors in the
derived mass profile.

In addition to the ``hot'' clusters ($T>5$~keV) just described,
interesting constraints on the mass profiles have also been
constrained for cooler clusters. The {\sl XMM} data of the cool
$(T\approx 2$~keV) cluster A1983 have been analyzed by
\citeN{prat03a}. The X-ray image and radial surface brightness
profiles are quite regular. The temperature profile of the hot gas is
well constrained (and slightly rising) out to $0.5$~Mpc.  Pratt \&
Arnaud do find that the NFW profile is a good fit to the data, but the
inferred concentration parameter $(c=3.8\pm 0.7)$ is lower than
expected ($\approx 9$) for this low-mass cluster. They argue that
departures from hydrostatic equilibrium are likely the cause of the
discrepancy, however, the regularity of the X-ray image does not
provide obvious support for that possibility. The authors also find
marginal evidence for multi-temperature gas in the central region
which should also be considered a possible origin for the anomalous
concentration value.

Many more clusters also have been analyzed with {\sl Chandra} and {\sl
XMM} observations; e.g., A2199
\cite{john02a}, 3C295 \cite{alle01b}, and MS1008.1-1224
\cite{etto03a}. Overall, NFW (and Moore et al.) profiles are found to be good
fits to the mass profiles outside cluster cores ($0.3-0.7$~$r_{\rm
vir}$) obtained from {\sl Chandra} and {\sl XMM}
observations. Exceptions (particularly A1983) are probably the result
of departures from hydrostatic equilibrium and/or multi-temperature
gas.

\subsection{Inside the Core}

\begin{figure}
\plottwo{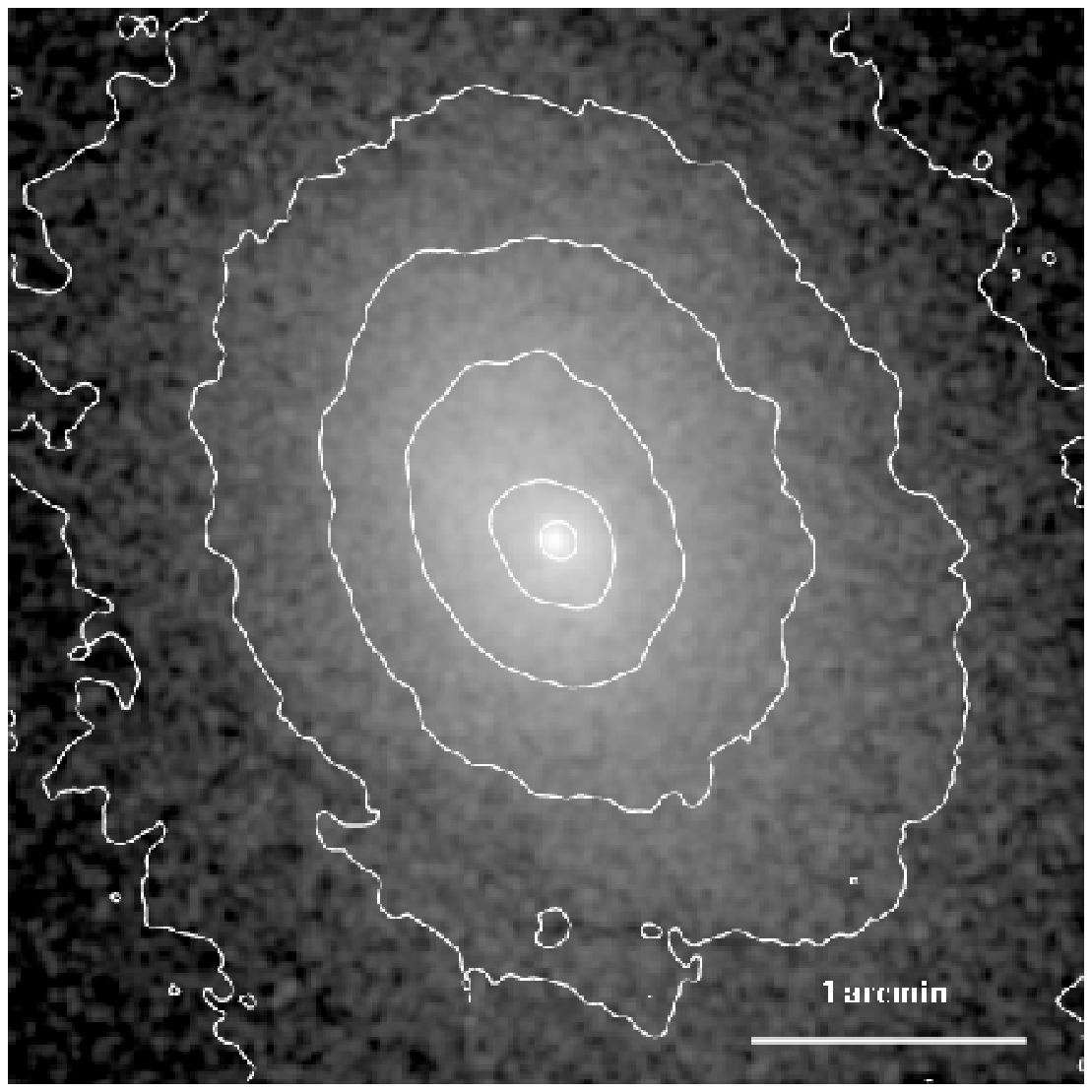}{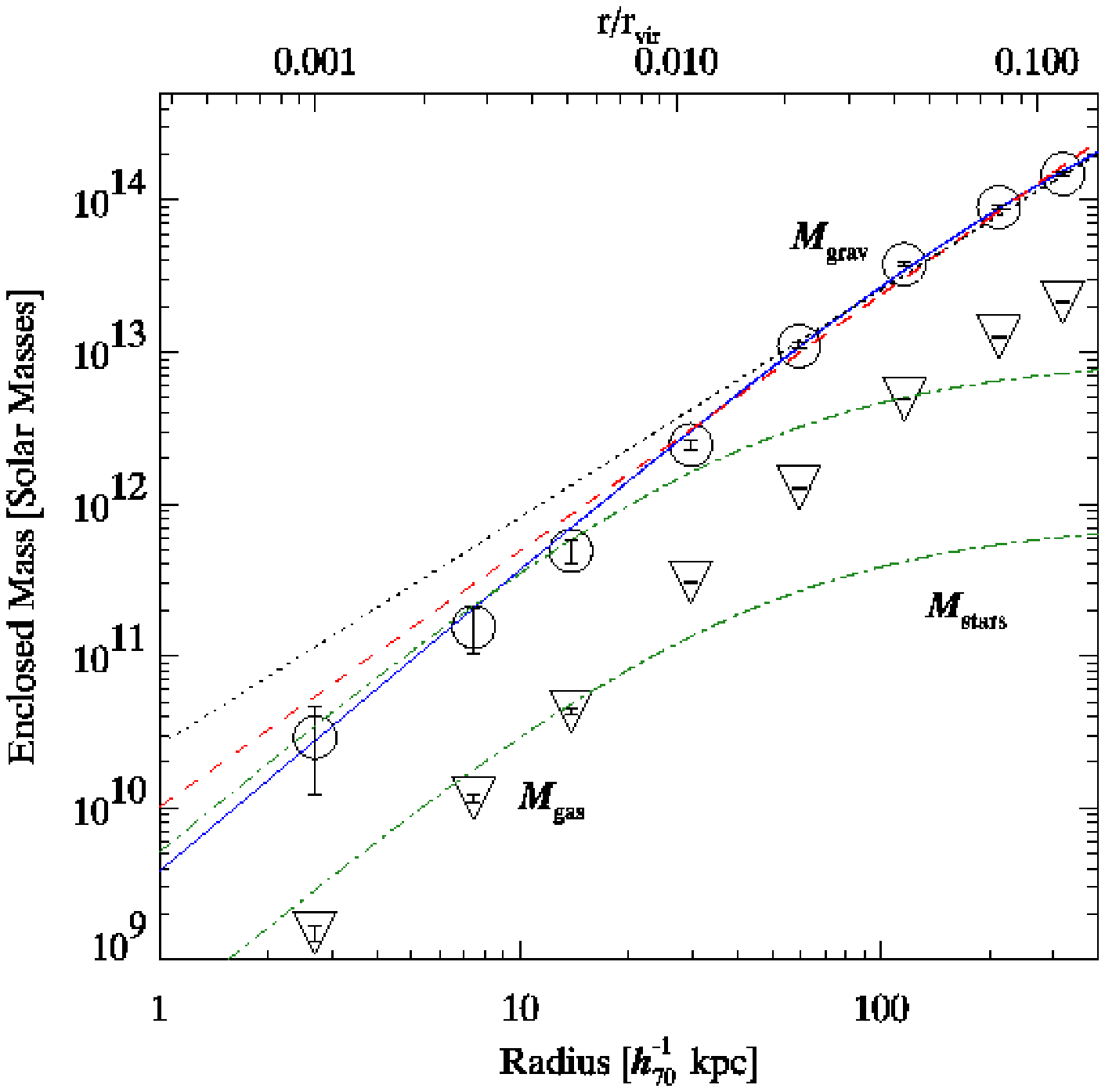}
\caption{\label{fig.a2029} From Lewis et al.\ (2002,2003): ({\it Left Panel}) {\sl
Chandra} ACIS-S image of Abell 2029. The image is 4$\arcmin$
($348h_{70}^{-1}$~kpc) on a side and has been smoothed with a Gaussian
of $\sigma = 2\arcsec$. ({\it Right Panel}) Total gravitating mass
(data points enclosed with open circles), overlaid with three
different mass models: NFW (\emph{solid curve}), power-law
(\emph{dashed line}), and Moore et al.\ (\emph{dotted curve}). The gas
mass is plotted as data points enclosed with open triangles.  Stellar
mass (\emph{dot-dashed curves}): lower curve assumes a $M_*/L_V$ of 1,
the upper curve 12.  Note: $r_{\rm vir}=2.71h_{70}^{-1}$~Mpc.}
\end{figure}

CDM predicts that DM halos should have approximately a power-law form
at small radius $(\rho(r)\propto r^{-\alpha})$, though the precise
value of the power-law exponent remains controversial. For CDM,
simulations indicate that $\alpha$ is between 1 (NFW) and 1.5 (Moore
et al.). But observations of LSB galaxies indicate $\alpha\approx 0.5$
in disagreement with CDM. This discrepancy is the motivation of the
``self-interacting'' DM model (SIDM) proposed by \citeN{sper00}. It is
natural to ask whether clusters also have $\alpha\approx 0.5$ since
they are DM-dominated deep into their cores. Of course, to address
this issue we need to focus on clusters with relaxed cores where the
assumption of hydrostatic equilibrium will be valid.

Unfortunately, {\sl Chandra} observations have shown that nature has
provided a very serious roadblock to using X-ray observations to study
the mass profiles in cluster cores. Clusters that tend to be the most
relaxed systems on hundred-kpc and Mpc scales have usually been
associated with ``cooling flows''. But observations with {\sl Chandra}
have shown that in the central regions of cooling flows (i.e., within
$r\approx 50$~kpc) the X-ray surface brightness is highly
disturbed. The hot gas exhibits filaments and holes -- where these
holes are often filled by radio emission from an AGN in the central
galaxy. Such morphological irregularities clearly call into question
the assumption of hydrostatic equilibrium. 

Attempts to measure the core mass profile using {\sl Chandra}
observations of clusters with such obvious disturbances have found
disagreement with CDM. Although \citeN{davi01a} obtain $\alpha\approx
1.3$ in the Hydra-A cluster, they obtain a concentration parameter
$(c=12.3\pm 0.2)$ that is higher than expected. In the core of A1795
\citeN{etto02a} find $\alpha=0.59\pm 0.15$ in clear disagreement with
CDM. 

It is difficult to evaluate these results for Hydra-A and A1795 since
departures from hydrostatic equilibrium could easily explain the
discrepancies with CDM. Clearly, it is essential to find X-ray clusters
that are relaxed in their cores. Some such clusters have now been
identified by considering those systems that are known to be be relaxed on 
hundred-kpc and Mpc scales but also do not have a strong radio source
in the central galaxy. 

One such cluster is A2029 (Figure \ref{fig.a2029}) which {\sl Chandra}
shows is indeed regularly shaped all the way down into its core
\cite{lewi02a,lewi03a}. This massive cluster is also nearby
($z=0.0767$) and very bright, making it an ideal target for X-ray
analysis of its core mass distribution. The {\sl Chandra} spectra of
A2029 extracted within thin annuli from the center out to $r\approx
300$~kpc are well-fitted by a single temperature component. This is a
key result since one may have expected possible complications from a
multiphase gas in the center of a cooling flow. 

These authors find that the mass profile of A2029 is reasonably well
described by a simple power law that is quite insensitive to different
parameterizations of the measured gas density and temperature
profiles. The exponent of the power-law fit to the gravitating mass
profile is $1.68\pm 0.02$. This translates to $\alpha=1.32\pm 0.02$
for the mass density profile, very similar to the NFW/Moore values,
yet very significantly larger than the values $\alpha\approx 0.5$
obtained for LSB galaxies. The NFW model is a better fit to the mass
profile than either a power-law or the Moore et al.\ model -- though a
fit of similar quality is obtained for the Hernquist profile. The
inferred concentration parameter is reasonable $(c=5.9\pm 0.3)$. After
subtracting out the contribution to the gravitating mass from the hot
gas and stars, it is found that the DM has very similar properties to
that of the gravitating matter: $\alpha=1.25\pm 0.03$ for a power-law
mass model and $c=4.6\pm 0.4$ for NFW. (Note: these results quoted for
A2029 include an extra data point at large radius not used by
\shortciteNP{lewi03a} and thus differ slightly from the previously
published results.)

In \citeN{buot03d} we present a {\sl Chandra} observation of A2589
which is another cluster that is found to be regularly shaped all the
way into into its core. Our {\sl Chandra} observation unfortunately
only yielded 8~ks of good data, and thus the constraints are not as
precise as those obtained for A2029. Like A2029, the hot gas in A2589
is consistent with a single-phase gas. The gas density and temperature
are well constrained and described by simple functions -- the gas is
consistent with being isothermal. The results for the gravitating mass
profile are quite consistent with those obtained for A2029. A
power-law fit to the mass profile yields $\alpha=1.37\pm 0.14$ for the
mass density. A better fit is provided by NFW which is a good fit down
to $\approx 0.01r_{\rm vir}$ with a reasonable concentration parameter
$(c=6.6\pm 2.5)$. Similar results are obtained for the DM:
$\alpha=1.35\pm 0.21$ for a power-law mass model and $c=4.9\pm 2.4$
for NFW.

Similar agreement with CDM is also observed for some clusters
\cite{alle02b,arab02a} that are not quite so obviously relaxed in
their cores as A2029 and A2589 but are clearly not as disturbed as
systems like Hydra-A and A1795.  Hence, in contrast to the clusters
with unrelaxed cores, the core DM profiles inferred for clusters with
relaxed cores appear to be quite consistent with CDM.

\section{Shapes of DM Halos in Elliptical Galaxies}

\begin{figure}
\plottwo{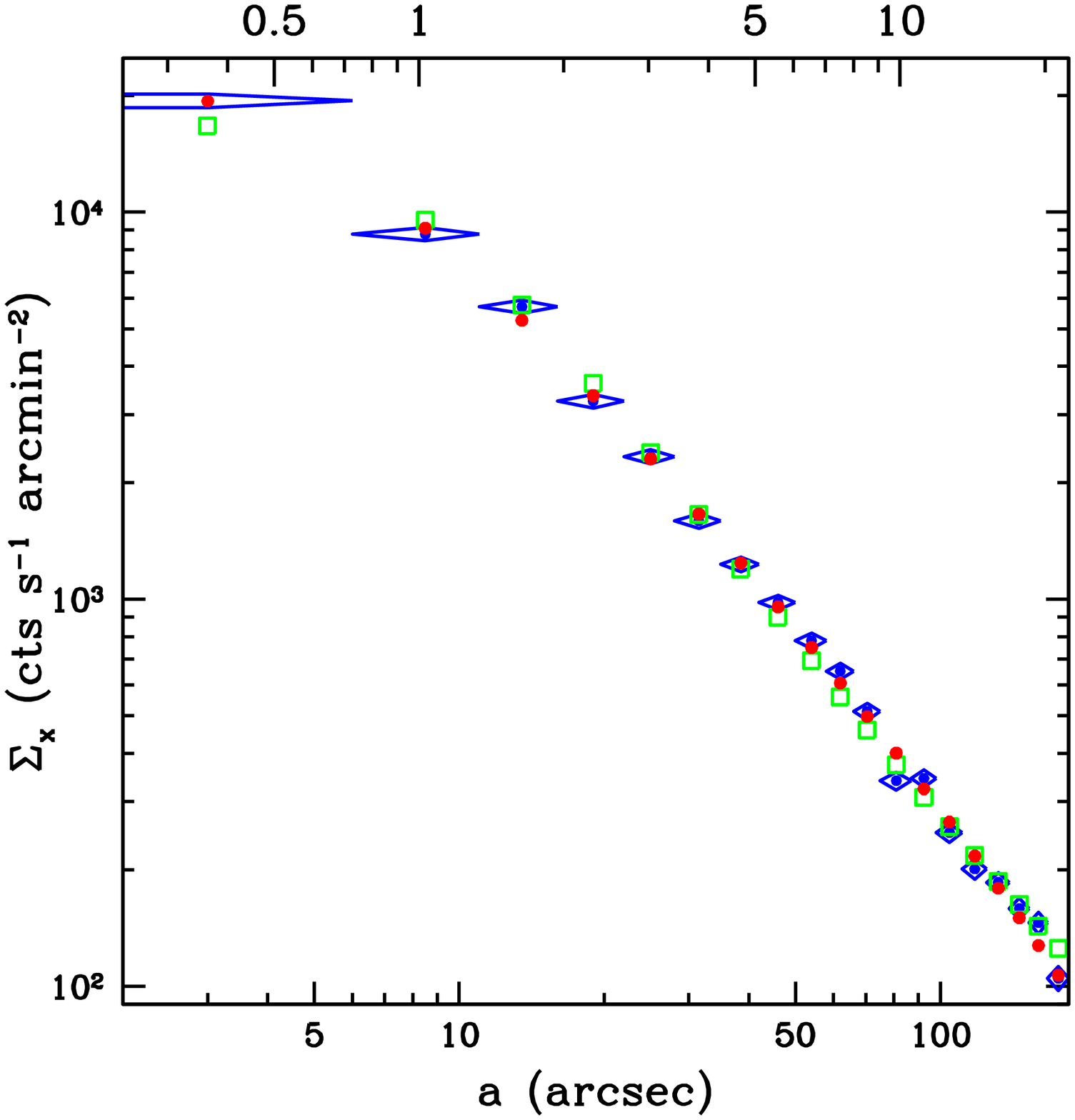}{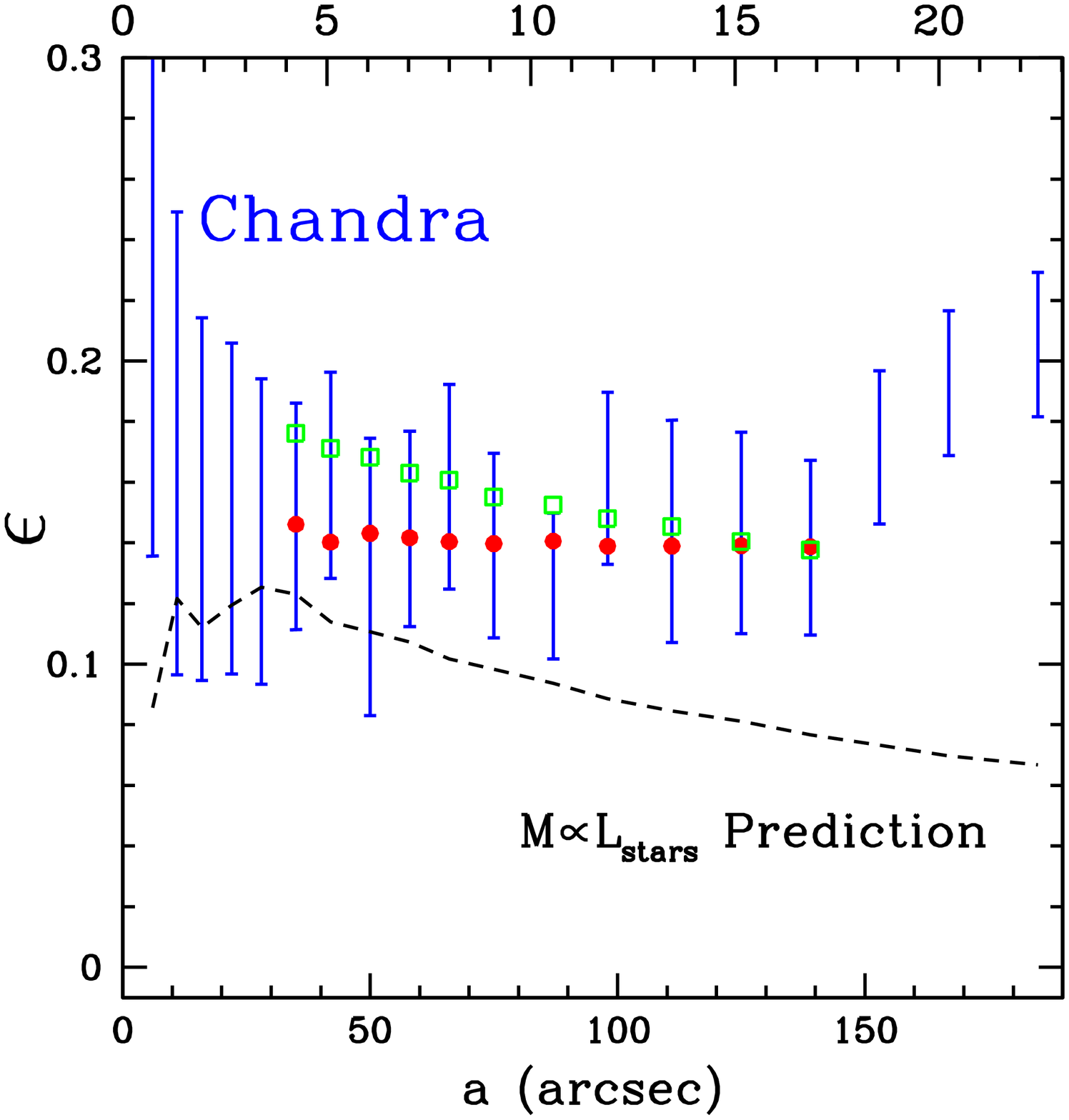}
\caption{\label{fig.n720} From Buote et al.'s (2002) analysis of the
{\sl Chandra} data 
of the elliptical galaxy NGC~720. ({\sl Left panel}) Radial surface
brightness profile denoted by diamonds (blue). Also shown are the
best-fitting radial profiles generated by (oblate) DM halos
corresponding to a $\rho\sim a^{-2}$ profile (filled circles -- red)
and an NFW profile (open squares -- green). ({\sl Right panel}) X-ray
ellipticities predicted by these DM models. The error bars (blue) are
the values of $\epsilon_x$ measured from the source-free {\sl chandra}
data, and the dashed line is the prediction if mass follows the
stars. We express $a$ in kpc on the top axis.}
\end{figure}

Unfortunately, since the X-ray halos of individual elliptical galaxies
are much fainter than those of clusters and groups of galaxies, it is
much more difficult to obtain interesting constraints on the
gravitating mass in elliptical galaxies using X-rays.  In fact, only
for NGC 720 has a detailed analysis of the mass distribution using
{\sl Chandra} data been published so far. 

The ellipticities of DM halos provide important constraints on the
nature of the DM in cosmological models essentially independent of
$\Omega_m$ or $\Lambda$
\cite{fren88a,bull02a}.  Nearly spherical halos are expected
for the cores of DM halos if the dark matter is self-interacting
\cite{dave01a}. Highly flattened (nearly disk-like) DM halos
occur in models where the DM is in the form of cold, molecular gas
\cite{pfen94}. Halos formed in a $\Lambda$CDM Universe are
moderately flattened with typical ellipticities $0.3-0.5$, though they
are rounder if there is significant ``warm'' DM
\cite{bull02a}.

In contrast to the case of spherical symmetry discussed so far, there
is no simple analytic solution of the equation of hydrostatic
equilibrium for ellipsoidal mass distributions. However, the added
complexity of ellipsoidal symmetry does offer some subtle rewards. For
a single-phase gas in hydrostatic equilibrium the volume X-ray
emissivity and the gravitational potential have identical shapes
independent of the temperature profile (An ``X-ray Shape Theorem'' --
\shortciteNP{buot94,buot96a}). This allows a robust ``Geometric Test'' of
the hypothesis that mass follows the optical light. One just
deprojects the stellar image and computes the potential assuming mass
follows the optical light. The shape of the potential is then compared
to the shape of the deprojected X-ray image independent of the gas
temperature profile.

This test for DM has been applied to a {\sl Chandra} observation of
the nearby ($D=25$~Mpc) galaxy NGC 720 \cite{buot02b}. This is an
isolated E4 galaxy that is fairly bright in X-rays. The {\sl Chandra}
image of this galaxy revealed over 40 discrete sources embedded within
the diffuse emission within $2.5\arcmin$ of the galaxy center. These
sources, most of which were undetected by previous observations, are
wonderful for studying the properties of X-ray binaries, but are a
nuisance for studies of DM.  These sources were removed from the image
and replaced with estimates of the local diffuse emission. The image
of NGC 720 cleaned of discrete sources displays X-ray isophotes with
moderate flattening and orientations that are slightly offset from the
optical.

Starting at the center of the galaxy, the positional angles (PAs) of
the isophotes slowly twist $\approx 20\deg$ away from the optical
major axis out to $r\approx 15$~kpc. There is a further twist at
larger radii. Here we focus on the region within $\approx 15$~kpc. If
the gas is in hydrostatic equilibrium, since this twist does not occur
in the stellar image it implies the existence of a massive DM halo
according to the Geometric Test mentioned above.

The X-ray ellipticity profile is consistent with a constant value
($\approx 0.15$) out to $\approx 15$~kpc at which point it
increases. Again, we restrict our discussion to the region within this
radius. The X-ray ellipticities are much less than the stellar values
($\approx 0.4$). But they are larger than predicted by the Geometric
Test (Figure \ref{fig.n720}). The mass-follows-optical-light
hypothesis is rejected -- using ellipticities alone -- at slightly
more than the $2\sigma$ level.

The evidence for DM provided by the X-ray PA twist and ellipticities
cannot be explained away by MOND. First the need for DM occurs already
in the region where Newtonian gravity should dominate. Second, the
potential shapes in MOND are the same as in Newtonian theory. Note
that gas rotation, which is not accounted for in this analysis, is not
expected to be important for NGC 720 \cite{buot02b}.

We find that the ellipticity of the gravitating mass distribution,
which is dominated by the DM, is not very sensitive to the mass model
(Figure \ref{fig.n720}). We find $\epsilon=0.37\pm 0.03$ for $\rho\sim
a^{-2}$ and $\epsilon=0.41\pm 0.02$ for NFW. Note that if the stars
make a sizable contribution to the potential (which is not supported
by the {\sl Chandra} data) then the ellipticity of the DM will be
larger than that of the gravitating matter.

The DM ellipticity inferred for NGC 720 is consistent with CDM
predictions but is very inconsistent with the round halos predicted by
SIDM and the very flat halos predicted if the DM is cold gas. The PA
twist, particularly within $\approx 15$~kpc strongly suggests the DM
is triaxial -- the twist appears to agree with the triaxial model of
NGC 720 proposed by \citeN{roma98}.

Although these results give nice agreement with CDM predictions, it is
essential to verify them using independent methods and with other
galaxies. The stellar velocity field could offer such an independent
test since the strongly triaxial mass distribution indicated by the
X-ray data should produce a highly asymmetric stellar velocity field
(T. Statler, private communication). We are currently analyzing new
observations of the elliptical galaxies NGC 1332 and NGC 3923 which
will provide key comparisons to mass properties of NGC 720.

We emphasize that NGC 720 shows the importance of detecting and
accounting for discrete sources embedded in the diffuse
emission. \citeN{stat02a} argue that the {\sl Chandra} measurements of
highly flattened X-ray isophotes of the elliptical galaxy NGC 1700
indicate that rotation must be dynamically important in the hot
gas. However, these authors detect only 2-3 discrete sources in their
system, and the X-ray ellipticity profile they derive matches the
optical ellipticities over almost the whole radius range
investigated. This strongly suggests that that the morphology of the
X-ray emission of NGC 1700 is strongly biased by the unresolved
discrete sources.

\section{Conclusions}

Presently, the key result obtained from {\sl Chandra} and {\sl XMM}
studies of DM in clusters and ellipticals is that the radial density
profiles in galaxy clusters and the ellipticity of the DM halo (for
NGC 720) agree well with CDM predictions. In particular, the DM
profiles measured deep down into the cores of the relaxed galaxy
clusters A2029 and A2589 rule out an important contribution from
self-interacting DM of the kind proposed to account for DM in the
cores of LSB galaxies.

\acknowledgements I thank the conference organizers for the invitation
to provide this review and for arranging such an excellent meeting. I
also thank my collaborators on X-ray studies of DM for their
contributions to some of the papers discussed in this review.

\bibliographystyle{pasp}

\end{document}